\documentclass[aps,prl,twocolumn,showpacs,showkeys,nofootinbib,superscriptaddress]{revtex4-1}

\usepackage[utf8]{inputenc}
\usepackage{graphicx}
\usepackage{amsfonts}
\usepackage{amssymb}
\usepackage{amsmath}
\usepackage[mathscr]{eucal}
\usepackage{setspace}
\usepackage{booktabs}
\usepackage[shellescape,latex]{gmp}
\usempxpackage{amssymb}

\def\bb{\mathbb}
\def\sqr#1#2{{\vcenter{\hrule height.#2pt
   \hbox{\vrule width.#2pt height#1pt \kern#1pt
      \vrule width.#2pt}
   \hrule height.#2pt}}}

\def\bsqr#1#2{{\vrule width #1pt height#2pt}}
\def\bsquare{{\mathchoice\bsqr66\bsqr66\bsqr33\bsqr33}}
\def\badbreak{\penalty1000}

\def\R{{\bb R}}				    
\newcommand{\cI}{{\cal I}}                  
\newcommand{\cO}{{\cal O}}                  
\newcommand{\cS}{{\cal S}}                  
\newcommand{\cC}{{\cal C}}                  
\def\fir{{\scriptscriptstyle{\text{\rm IR}}}}             
\def\fuv{{\scriptscriptstyle{\text{\rm UV}}}}             
\def\lm0{{\lambda_0}}                                     
\def\nrN{N}                                               
\def\cf{\mathfrak{n}}                                     
\def\cd{\mathfrak{g}}                                     
\def\cfu{\cf_\star}                                       
\def\efN{\mathscr{N}}                                     
\def\efNm{\efN_\star}                                     
\def\Obj{O}                                                     
\def\Osec{\{\Obj\}}                                            
\def\esup{\Obj_{\text{s}}}                                
\def\w{c}                                                         
\def\W{C}                                                        
\def\Wsec{\{\W\}}                                             
\def\v{b}                                                            
\def\wl{w}                                                          
\def\Wl{W}                                                         

\def\Nmaps{{\mathfrak N}}                                    
\def\Nmapssup{\Nmaps_{\text{s}}}                       

\def\dmink{d_{\text{M}}}                                           

\makeatletter
\newcommand*{\smallrel}[2][.8]{%
  \mathrel{\mathpalette{\smallrel@{#1}}{#2}}%
}
\newcommand*{\smallrel@}[3]{%
  \sbox0{$#2\vcenter{}$}%
  \dimen@=\ht0 %
  \raise\dimen@\hbox{%
    \scalebox{#1}{%
      \raise-\dimen@\hbox{$#2#3\m@th$}%
    }%
  }%
}
\makeatother

\usepackage{amsmath} 
\usepackage{dsfont}  
\usepackage{bm}      

\def\beq{\begin{equation}}
\def\eeq{\end{equation}}
\def\beqs#1\eeqs{\beq\begin{split} #1 \end{split}\eeq}

\long\def\comment#1{}

\def\be{\begin{equation}}
\def\ee{\end{equation}}

\def\bc{\begin{center}}
\def\ec{\end{center}}

\usepackage{microtype}
\usepackage[dvipsnames]{xcolor}
\usepackage[colorlinks=true,backref=false, linktocpage=true,
citecolor=PineGreen,urlcolor=PineGreen,linkcolor=PineGreen,pdfpagemode=UseOutlines]{hyperref}

\hypersetup{%
  bookmarksnumbered=true,
  pdftitle = {},
  pdfsubject = {},
  pdfauthor = {},
  pdfkeywords = {}
}

\begin{document}

\title{Counting-Based Effective Dimension and Discrete Regularizations}

\author{Ivan Horv\'ath}
\email{ihorv2@g.uky.edu}
\affiliation{Nuclear Physics Institute CAS, 25068 \v{R}e\v{z} (Prague), Czech Republic}
\affiliation{University of Kentucky, Lexington, KY 40506, USA}

\author{Peter\ Marko\v s}
\email{peter.markos@fmph.uniba.sk}
\affiliation{Dept. of Experimental Physics, Faculty of Mathematics, 
Physics and Informatics, Comenius University in Bratislava, Mlynsk\'a Dolina 2, 
842 28 Bratislava, Slovakia}

\author{Robert Mendris}
\email{rmendris@shawnee.edu}
\affiliation{Shawnee State University, Portsmouth, OH 45662, USA}

\date{Mar 10, 2023}

\begin{abstract}

Fractal-like structures of varying complexity are common in nature, and 
measure-based dimensions (Minkowski, Hausdorff) supply their basic geometric 
characterization. However, at the level of fundamental dynamics, which is quantum, 
structure does not enter via geometry of fixed sets but is encoded 
in probability distributions on associated spaces. The question then arises whether 
a robust notion of fractal measure-based dimension exists for structures 
represented in this way. 
Starting from effective number theory, we construct all counting-based 
schemes to select effective supports on collections of objects with probabilities 
and associate the {\em effective counting dimension} (ECD) with each. 
We then show that ECD is scheme-independent and, thus, a well-defined 
measure-based dimension with meaning analogous to the Minkowski dimension 
of fixed sets. In physics language, ECD characterizes probabilistic descriptions
arising in a theory or model via discrete ``regularization''. For example, our 
analysis makes recent surprising results on effective spatial dimensions in 
quantum chromodynamics and Anderson models well founded. We discuss how 
to assess the reliability of regularization removals in practice and perform 
such analysis in the context of 3d Anderson criticality.


\end{abstract}

\maketitle

{\bf 1.~Prologue. $\!$}
Consider the prototypical example of fractal structure, the ternary Cantor set 
$\cC \!\subset\! [0,1]$. Evaluating its Minkowski 
dimension~\cite{falconer2014fractal} involves introduction of the regularization 
parameter $a \!>\! 0$, namely the size of elementary interval (``box"), 
and the use of ordinary counting to determine the number $\nrN(a)$ of such 
boxes required to cover $\cC$. The scaling of $\nrN(a)$ in the process 
of ``regularization removal", namely 
\begin{equation}
   \nrN(a) \,\propto\, a^{-\dmink} \quad \text{for} \quad  a \to 0
   \label{eq:001}                     
\end{equation}   
then specifies the Minkowski dimension $\dmink[\cC]$ ($=\!\log_3 2$).

Assume now that, instead of a fixed set such as $\cC$, we are given 
a probability measure $\mu$ over the sample space $[0,1]$ as a way
to introduce structure on this interval. The ensuing probabilities make 
certain parts of the interval preferred to others, which is the probabilistic 
analogue of sharply selecting $\cC$ in case of fixed sets. Is it possible 
to characterize the probabilistic case by a robust fractal dimension with 
meaning akin to Minkowski? 

Here we construct such dimension and clarify in what sense it is unique. 
To convey the idea, consider a schematic analogue of Minkowski prescription
in the above example. As a {\bf first step}, introduce a discrete regularization 
parameter $a \!=\! 1/\nrN$, where $\nrN$ now refers to number of equal-size 
intervals forming a partition $\cI \!=\! \{ \cI_i \,;\, i\!=\!1,2,\ldots , \nrN\}$ of 
$[0,1]$. With each $\cI_i$ associate the probability $p_i \!=\! \mu[\cI_i]$ to 
obtain the distribution $P(a) \!=\! (p_1,\ldots, p_{\nrN(a)})$. Thus, for each 
$a \!\in\! \{1,1/2,1/3,\ldots\}$ we have a collection of Minkowski~boxes 
which, however, come with probabilities.
In a {\bf second step}, assume that we modify ordinary counting 
$\nrN \!=\! \nrN[\cI]$ of boxes to $\efN \!=\! \efN[\cI,P]$ so that 
probabilities are properly taken into account. In fact, $\efN[\cI,P] \!=\! \efN[P]$ 
since $P$ already carries the information on $\nrN[\cI]$. 
Scaling of $\efN$ upon the regularization removal, namely
\begin{equation}
   \efN[P(a)] \,\propto\, a^{-d_\fuv} \quad \text{for} \quad a \to 0
   \label{eq:002}                     
\end{equation}   
would then specify the dimension $d_\fuv \!=\! d_\fuv[\mu,\efN]$ in analogy 
to \eqref{eq:001}. The subscript UV (ultraviolet) conveys that regularization 
controls the structure at small distances. In this sense, $\dmink$ is of course 
a UV construct as well.

The above plan for $d_\fuv$ becomes a well-founded concept analogous 
to $\dmink$ assuming that: 
{\em (i)} scheme $\efN$ is additive like ordinary $\nrN$;
{\em (ii)} there is a notion of well-delineated effective support 
$\cS \!=\! \cS[\mu,\efN]$ induced by $\mu$ and specified by $\efN$, 
namely an analog of $\cC$;
{\em (iii)} $d_\fuv$ only depends on $\mu$, i.e.
$d_\fuv[\mu,\efN] \!=\! d_\fuv[\mu]$ for all $\efN$ satisfying
{\em (i)} and~{\em (ii)}. 
Indeed, {\em (i)} makes $d_\fuv$ measure-based in the same sense 
$\dmink$ is, and {\em (ii)} allows for a structure induced probabilistically by $\mu$ 
to be described in the same way as structure of fixed sets. Part {\em (iii)} then
guarantees that $d_\fuv$ is robust (unique).

The notion of effective counting is clearly central to~$d_\fuv$. Its theoretical
framework is the effective number theory (ENT)~\cite{Horvath:2018aap} 
which, among other things, determines all $\efN$ satisfying {\em (i)}. 
In this work we give an affirmative status to {\em (ii)} and {\em (iii)} by
developing the concept of {\em effective counting dimension} 
(ECD), which is more general than~$d_\fuv$.

Before giving a self-contained account of ECD, few points are 
worth emphasizing. {\em (a)} The variety of structures characterized 
by $d_\fuv$ is much larger than~that by $\dmink$. For example, 
$\mu$ may describe Cantor set with non-uniform probability measure 
on it, for which $\dmink$ is not applicable. 
{\em (b)} While probability/stochasticity play important role in the fractal 
world~\cite{bishop_peres_2016}, 
they usually enter differently than here. For example, representative 
of a random Cantor set or an individual Brownian path arise via 
a random process, but such sample is still a fixed set. Our extension 
in these situations treats sample itself as a probability~measure. 
{\em (c)} The above setup arises e.g. in quantum description of natural 
world since quantum states encode probabilities of physical events. 
Here it is common that dynamics is in fact defined by regularization, 
be it UV or IR (system size $L \!\to\! \infty$). The input for computing 
$d_\fuv$ or $d_\fir$ is then directly $P(a)$ or $P(L)$ rather than 
(usually unknown) $\mu$. 
Such calculations of $d_\fir$ for Dirac eigenmodes in quantum 
chromodynamics~\cite{Alexandru:2021pap} and for critical states of 
Anderson transitions~\cite{Horvath:2021zjk} recently led to new geometric 
insights.

\smallskip

\noindent
{\bf 2.~Qualitative Outline \& Summary. $\!$}
In the study of physical and various model situations we frequently need 
to describe or analyze some collection $\Obj \!=\! \{o_1,o_2, \ldots \}$ of 
objects. The most basic quantitative analysis of $\Obj$ is to count its 
objects, namely to label it by a natural number $\nrN$. 
However, descriptive value of an ordinary count becomes limited 
when objects in a collection differ substantially. Indeed, consider
$\Obj_\bullet$ containing the Sun, the Proxima Centauri and 
$10^{10}$ individual grains of sand from Earth's beaches. 
The ordinary count $\nrN \!=\! 2 + 10^{10}$ is clearly a poor 
characteristic of $\Obj_\bullet$ if the individual importance of 
its objects is judged by their masses.

The recent 
works~\cite{Horvath:2018aap,Horvath:2018xgx,Horvath:2019qeo} 
revisited counting with the aim of making it more informative in such 
situations. The resulting effective number theory studies possible 
ways to assign counts to collections of
objects distinguished by an additive importance weight such as 
probability or mass. For example, with $\Obj_\bullet$ it associates 
an {\em intrinsic (minimal) effective count} by mass 
$\efNm \!=\! 2 + 2.0 \times 10^{-16}$, leading to useful 
quantitative insight. [The textbook masses for the stars and the average 
mass of $4.5\,$mg for a grain of sand went into this calculation 
using Eq.~\eqref{eq:040}. Note that a specific normalization is 
involved in the prescription.]

The utility of ordinary counting in analyzing the natural world 
stems mainly from its {\em additivity} which guarantees consistent 
bookkeeping for stable objects and leads to predictability: given 
disjoint collections $\Obj_1$ and $\Obj_2$ with labels $\nrN_1$ 
and $\nrN_2$ we can predict what the label of the combined 
collection $\Obj_1 \cup \Obj_2$ would be ($\nrN_1 \!+\! \nrN_2$) 
without actually performing the merger. Since merging and 
partitioning are at the heart of dealing with objects, counting 
became a theoretical tool: it is usually easier and faster 
to handle numbers than physical objects.

ENT requires any extension of counting to its effective form to be 
additive in order to preserve the above features~\cite{Horvath:2018aap}.
Generalized additivity arises naturally if one views ordinary counting 
as a process carried out by a machine. Its ``ready state" includes
an empty list $I\!=\!()$ and its operation entails receiving objects 
from $\Obj$ sequentially. Each input causes $I$ to update via 
$I \rightarrow I \sqcup (1)$, where $\sqcup$ 
denotes concatenation, resulting in a certain $I=(1,1,\ldots,1)$ upon 
exhausting all objects. The machine then calls its number function 
$\nrN \!=\! \nrN[I]$ (length of a list) and outputs the label. This 
representation of ordinary counting makes it plain that the scheme 
is encoded by function $\nrN[I]$ and its additivity is expressed by 
the functional equation 
\vspace{-0.04in}
\begin{equation}
    \nrN[I_1 \sqcup I_2] \!=\! \nrN[I_1] + \nrN[I_2]
    \quad , \quad \forall \,I_1 \,,\,I_2
     \label{eq:010}                 
\end{equation}

\vspace{-0.1in}
In effective counting, each object $o$ comes with a label specifying 
its own weight $\wl$. The associated machine initializes the weight list 
$\Wl\!=\!()$ and then sequentially inputs the objects. Upon each input, 
it scans the weight $\wl$ and updates $\Wl$ via $\Wl \rightarrow \Wl \sqcup (\wl)$. 
After finishing the input, it calls its number function $\nrN \!=\! \nrN[\Wl]$ 
to get the ordinary count, and uses it to rescale $\Wl$ into a canonical counting 
form $\Wl \!\to\! \W=(\w_1,\w_2,\ldots,\w_\nrN)$ satisfying 
$\sum_i \w_i \!=\! \nrN$. This rescaling is allowed since, like 
its ordinary prototype, effective counting is scale invariant by construction.
The machine then calls its effective number function $\efN \!=\! \efN[\W]$ 
and outputs the result.  Additivity of the procedure is then expressed 
by the functional equation~\cite{Horvath:2018aap}
\begin{equation}
   \efN[\W_1 \sqcup \W_2] = \efN[\W_1] + \efN[\W_2] 
   \quad , \quad \forall \,\W_1 \,,\,\W_2
   \tag{A}
   \label{eq:add}                   
\end{equation}
It ensures that effective counts of disjoint collections with 
equal average weight per object add up upon merging.

Hence, in the same way that $\nrN \!=\! \nrN[I]$ encodes ordinary 
counting, each $\efN \!=\! \efN[\W]$ obeying \eqref{eq:add} and 
other necessary conditions~\cite{Horvath:2018aap} specifies 
a valid effective counting scheme. When used consistently
in all situations, each offers bookkeeping and predictability features 
analogous to those of ordinary counting. 
ENT identifies all such effective schemes $\efN$. A key property of 
the resulting concept is that the scheme specified by~\cite{Horvath:2018aap}
\begin{equation}
      \efNm[\W]  \,=\, \sum_{i=1}^\nrN \cfu(\w_i)   \quad,\quad
      \cfu(\w)  \; = \;   \min\, \{ \w, 1 \}    \;
      \label{eq:040}         
\end{equation}
satisfies $\efNm[\W] \!\le\! \efN[\W] \!\le\! \nrN[\W]$ for all 
$\W$ and all $\efN$. Hence, the effective total prescribed by $\efNm$ 
cannot be lowered by a change of counting scheme and is intrinsic 
to a collection. In fact, each 
collection of objects with additive weights is characterized by two 
key counting characteristics: the ordinary count $\nrN$ and 
the intrinsic effective count $\efNm$. 

In this work we show that effective counting entails a unique notion of 
dimension. The associated setting involves an infinite sequence of collections 
$\Obj_k$ with $\nrN_k$ objects, and an associated sequence 
$\W_k \!=\! (\w_{\scriptscriptstyle{k,1}}, \w_{\scriptscriptstyle{k,2}}, 
\ldots, \w_{\scriptscriptstyle{k,\nrN_k}})$ of counting weights.
Here $\nrN_k$ is strictly increasing and hence 
$\lim_{k \to \infty} \nrN_k \!=\! \infty$. The pair $\Obj_k$, $\W_k$ 
may specify e.g. an increasingly refined representation of a complex 
composite object or of a physical system with infinitely many parts. 
Following the standard physics language, we refer to it as 
``regularization" of the target $k \!\to\! \infty$~situation.

Assume that we fix a counting scheme $\efN$ and associate with 
each weighted collection $\Obj$, $\W$ its effective description
$\esup \!=\! \esup[\W,\efN]$, containing only $\efN[\W]$ 
highest-weighted objects of $\Obj$. To any regularization sequence 
$\Obj_k$, $\W_k$ this assigns a sequence of effective descriptions
$\Obj_{\text{s},k}$ yielding the effective description of the target.
Then the idea of {\em effective counting dimension} (ECD) is to convey 
how the abundance of objects in the effective description of the target  
scales with that in its full representation. In other words, ECD corresponds 
to $\Delta$~in
\begin{equation}
   \efN[\W_k] \propto \nrN[\W_k]^{\,\Delta}
   \quad\text{for}\quad k \to \infty
   \quad\; , \quad\; 0 \le \Delta \le 1  \;
   \label{eq:060}
\end{equation}
However, it turns out that the above notion of effective description 
(effective support) $\esup$ is only consistent for certain schemes 
$\efN$. Indeed, for $\efN$ to delineate the support properly, a separation 
property formulated in Sec.~4 below has to be imposed.
Formally, if $\Nmaps$ is the set of all schemes $\efN$, then only 
elements of its subset $\Nmapssup \!\subset \Nmaps$ assign 
effective supports. We will show in Sec.~4 that $\Nmapssup$ 
is spanned by
\begin{equation}
   \efN_{(u)}[\W] \!=\! \sum_{i=1}^\nrN \cf_{(u)}(\w_i)
   \quad , \quad
   \cf_{(u)}(\w) = \min\, \{ \w/u, 1 \}  \;\;
   \label{eq:080}         
\end{equation}
where $u \in (0,1]$. Note that $\efN_{(1)} \!=\! \efNm \in \Nmapssup$.

The above leads us to consider $\Delta \!=\! \Delta[\Wsec,\efN\,]$
with $\Wsec$ a shorthand for regularization sequence and 
$\efN \in \Nmapssup$. The minimal nature of $\efNm$ implies 
that  $\Delta_\star[\Wsec] \equiv \Delta[\Wsec,\efNm]$ is the smallest 
possible ECD. However, ECD is in fact fully robust and doesn't 
depend on $\efN$ at all. Indeed, in Sec.~5 we will show that
\begin{equation}
   \Delta_\star[\Wsec] 
   \,=\,  \Delta[\Wsec,\efN\,]  \quad\; , \quad\;
   \forall \, \efN \in \Nmapssup   
   \label{eq:100}
\end{equation}
Hence, ECD is a well-defined property of the target specified by 
regularization pair $\Osec$, $\Wsec$. The use of additive 
counting makes it the measure-based effective dimension. 

Before demonstrating the results \eqref{eq:080} and \eqref{eq:100} 
we wish to make few remarks. 
{\em (i)} The fact that ECD doesn't require metric allows for large
range of applications. Indeed, models in some areas (e.g. ecosystems 
and social sciences) often do not involve distances.
{\em (ii)} Setups with metric frequently entail UV cutoff $a$ 
($\propto$ shortest distance) and IR cutoff $L$ 
($\propto$ longest distance). Sequence $\Wsec$ can 
facilitate their removals: $\W_k$ may be associated e.g. with 
$a_k \!\to\! 0$ at fixed $L$, or with $L_k \!\to\! \infty$ at fixed $a$. 
Defining the nominal dimensions via
$\nrN[\W_k] \!\propto a_k^{-D_{\text{UV}}(L)}$ and 
$\nrN[\W_k] \!\propto L_k^{D_{\text{IR}}(a)}$ 
for UV and IR cases, their effective counterparts are~\cite{Alexandru:2021pap}
\begin{equation}
   \efNm[\W_k] \propto a_k^{-d_{\text{UV}}(L)}  
   \quad\; , \quad\;
   \efNm[\W_k] \propto L_k^{d_{\text{IR}}(a)}   \quad
   \label{eq:110}   
\end{equation}
If $\Delta_\fuv$, $\Delta_\fir$ denote their associated ECDs then
\begin{equation}
   d_\fuv = \Delta_\fuv \, D_\fuv
   \quad\; , \quad\;
   d_\fir = \Delta_\fir \, D_\fir  \quad
   \label{eq:120}
\end{equation}
Dimension $d_\fir$ was recently calculated in QCD \cite{Alexandru:2021pap} 
and~in Anderson models~\cite{Horvath:2021zjk}.
{\em (iii)} The meaning of $d_\fuv$ is fully analogous to Minkowski 
(box-counting) dimension of fixed sets.
In fact, for $\Osec$ that UV-regularizes a bounded region in $\R^D$, 
$d_\fuv$ is exactly the Minkowski dimension of $\{ \Obj_s\}$ treated as 
a fixed set. Uniqueness \eqref{eq:100} of ECD suggests that 
measure-based dimensions are meaningful even for geometric figures 
emerging effectively, e.g. from probabilities. 
{\em (iv)} We refer to $\esup \!=\! \esup[\Obj,\W,\efN]$, 
$\efN \!\in\! \Nmapssup$, as both the support and the description of $\Obj$. 
The latter is more suitable in situations involving information and complexity. 
Our analysis implies the existence of well-defined 
{\em minimal effective description} 
$\Obj_\star[\Obj,\W]$, namely  $\esup[\Obj,\W,\efNm]$, which may find uses 
in these contexts. 

\smallskip

\noindent
{\bf 3.~Effective Counting Schemes. $\,$} Our starting point~is ENT 
\cite{Horvath:2018aap} which determines the set $\Nmaps$ 
of all effective counting schemes 
$\efN \!=\! \efN[\W] \!=\! \efN(\w_1,\ldots,\w_\nrN)$. Apart from 
additivity~\eqref{eq:add}, symmetry, continuity and boundary 
conditions, the axiomatic definition of $\Nmaps$ also ensures that 
increasing the cumulation of weights in $\W$ doesn't increase 
the effective number. This monotonicity is expressed by
\begin{equation}
      \efN(\ldots \w_i + \epsilon  \ldots  \w_j - \epsilon  \ldots)   \,\le\, 
      \efN(\ldots  \w_i  \ldots  \w_j  \ldots)   
      \quad   
      \tag{M}
      \label{eq:mon}
\end{equation}
for each $\w_i \!\ge\! \w_j$ and $0 \!\le\! \epsilon \!\le\! \w_j$.
The resulting $\Nmaps$ consists of additively separable functions 
$\efN(\w_1,\ldots,\w_\nrN) \!=\! \sum_{i=1}^\nrN \cf(c_i)$, such that 
the {\em counting function} $\cf \!=\!\cf(\w)$, $\w \in [0,\infty)$, is 
\vspace{-0.01in}
\begin{equation}
   \begin{aligned}[c]
   &(i) \; \text{continuous}\\
   &(ii) \,\text{concave}\\
   \end{aligned}
   \qquad\qquad
   \begin{aligned}[c]
   &(iii)\,\, \cf(0)=0\\
   &(iv)\;\, \cf(\w)=1 \;\,\text{for}\;\, \w\ge 1\\
   \end{aligned}
   \label{eq:160}
\end{equation}
Representation of $\efN \!\in\! \Nmaps$ by $\cf$ satisfying 
\eqref{eq:160} is unique. 

\smallskip

\noindent
{\bf 4. Effective Supports.} We will assume from now on that the order 
of objects in $\Obj =\{o_1,\ldots ,o_\nrN\}$ is set by their relevance in 
$\W$, i.e. that $\w_1 \ge \w_2 \ge \ldots \ge \w_\nrN$. 
Given a counting scheme $\efN$, we collect the first $\efN[\W]$ 
objects to form the intended effective support (effective description) 
$\esup$ of $\Obj$. Since $\efN$ is real-valued, we represent $\esup$ as 
\begin{equation}
   \Obj \!=\! \{o_1, \ldots ,o_\nrN\}    
   \;\longrightarrow\;\,
   \esup[\Obj,\W] = \{o_1, \ldots ,o_J, f\} 
   \label{eq:180}   
\end{equation}
where $J$ is the ceiling of $\efN[\W]$, and $\efN[\W] \!=\! (J \!-\! 1) \!+\! f$.
Hence, $0 \!<\! f \!\le\! 1$ is the fraction of $o_J$ included in $\esup$. 

The rationale for effective support so conceived is clear: 
$\esup$ is a subcollection of most relevant elements from $\Obj$ that 
behaves under the ordinary counting measure in the same way as $\Obj$ 
under the effective one. Indeed, additivity \eqref{eq:add} translates into
(dependence on $\efN$, $\W_1$, $\W_2$ is implicit) 
\begin{equation}
     \nrN\bigl[  \esup[ \Obj_1 \sqcup \Obj_2 ] \bigr]  
     =
     \nrN\bigl[ \esup[ \Obj_1 ] \bigr]  +  \nrN\bigl[ \esup[ \Obj_2 ] \bigr]   
     \label{eq:200}
\end{equation}
where an obvious real-valued extension of $\nrN[\ldots]$ to collections 
with fractional last element was made.

However, ENT axioms only deal with counting, and their compatibility
with the above notion of effective support needs to be examined. 
Effective numbers are crucially shaped by additivity \eqref{eq:add} and 
monotonicity \eqref{eq:mon}. With \eqref{eq:add} being the basis for 
$\esup$ via \eqref{eq:200}, it is \eqref{eq:mon} that requires attention.  
To that end, consider the operation on left-hand side of \eqref{eq:mon}, 
involving the last object included in $\esup$ (i.e. $o_J$) and first object 
fully left out ($o_{J+1}$). Since $o_J$ gains relevance at the expense of 
$o_{J+1}$, its presence in $\esup$ (measured by $f$) cannot 
decrease. This {\em separation property} \eqref{eq:ssp} is expressed by
\begin{equation}
      \efN(\ldots \w_J + \epsilon  ,  \w_{J+1} - \epsilon  \ldots)   \,\ge\, 
      \efN(\ldots  \w_J  ,  \w_{J+1}  \ldots)   
      \;\;  
      \tag{SP}
      \label{eq:ssp}                   
\end{equation}
for all $\W$ such that $c_J \!>\! c_{J+1} \!>\! 0$ and all sufficiently~small 
$\epsilon \!>\! 0$. Ensuring a meaningful split of effective support from 
the rest, \eqref{eq:ssp} has to hold in order to define $\esup$ consistently. 
Note that $J\!=\!\text{\tt ceil}(\efN[\W])$ depends on both~$\efN$~and~$\W$.


We now show that the only counting schemes $\efN$ compatible with 
\eqref{eq:ssp} are specified by Eq.~\eqref{eq:080}. To start, note that in order
to make the separation property compatible with monotonicity~\eqref{eq:mon},
we have to impose the equality sign in \eqref{eq:ssp}. In terms of counting
function $\cf$ of $\efN$ we~have
\begin{equation}
    \cf(\w_J + \epsilon) + \cf(\w_{J+1} - \epsilon) =
    \cf(\w_J) + \cf(\w_{J+1}) 
    \label{eq:220}
\end{equation}
for all $\W$ with $c_J \!>\! c_{J+1} \!>\! 0$ and all sufficiently 
small $\epsilon\!>\!0$.

Note next that properties \eqref{eq:160} of $\cf$ imply the existence 
of $0 \!<\! u \!\le\! 1$ such that $\cf(c) \!=\! 1$ for all $c \!\ge\! u$, and 
$0 \!<\! \cf(c) \!<\! 1$ for all $0 \!<\! c \!<\! u$. Given this $u=u[\cf]$,
the separation operation in \eqref{eq:220} cannot be performed 
when $\w_J \!\ge\! u$. Indeed, since each 
$\cf(\w_j)$ with $j \!\le\! J$ contributes unity to $\efN[C]$, we have
$J \!=\! \text{\tt ceil}(J \!+\! \sum_{i=J+1}^\nrN \w_i)$, 
leading to $\w_{J+1} \!=\! 0$. 
Hence, it is sufficient to consider~\eqref{eq:220} for $\W$ with 
$u \!>\! c_J \!>\! c_{J+1} \!>\! 0$. In this form it is readily satisfied by 
$\cf_{(u)}$ of Eq.~\eqref{eq:080} due to its linearity on $[0,u]$. 
Consequently, $\efN_{(u)} \!\in \Nmapssup$.


However, all other $\cf$ featuring the same $u$ violate~\eqref{eq:220}. 
To show that, consider $\nrN \!=\! 3$ vectors $\W \!=\!(3 \!-\!y\!-\!x,y,x)$ 
with $0 \!<\! x \!<\! y \!<\! u$ and $J\!=\!2$. Definition of $J$, namely 
$J \!-\!1 \!< \efN[C] \!\le\! J$, demands that $\cf(x) + \cf(y) \le 1$. 
We will specify $x$, $y$ that satisfy this, as well as $\epsilon_0 \!>\!0$ 
such that $\cf(x - \epsilon) + \cf(y + \epsilon) \!<  \cf(x) + \cf(y)$ for all
$0 \!<\! \epsilon \!<\! \epsilon_0$, thus failing~\eqref{eq:220}.
Since $\cf_{(u)}$ produces equality in this relation, we can  
proceed using $\cd(x) \!=\! \cf(x) \!-\! \cf_{(u)}(x)$ and
\begin{equation}
    \cd(x - \epsilon) + \cd(y + \epsilon) \!<  \cd(x) + \cd(y)   
    \;\; , \;\;  0 < \epsilon < \epsilon_0
    \label{eq:230}
\end{equation} 
From properties \eqref{eq:160} of $\cf$, $\cf_{(u)}$ and the explicit form 
of $\cf_{(u)}$ it follows that $\cd(x)$ is a continuous function satisfying: \\
(a) $\cd(0) \!=\! \cd(u) \!=\! 0$;
(b) $\cd(x) \!>\! 0$ for $0 \!<\! x \!<\! u$; 
(c) there are $0 \!<\! x_0 \!\le\! y_0 \!<\! u$
such that $\cd(x)$ is increasing on $[0,x_0]$ and decreasing on $[y_0,u]$.
Hence, any $0 \!<\! x \le x_0$, $y_0 \!<\! y \!<\! u$ and 
$0 \!<\! \epsilon_0 \!\le\! \min\{x, u-y\}$ form a triple satisfying \eqref{eq:230}.
Finally, for any $y$ chosen as above, we can select sufficiently small 
$x \!>\!0$ such that $\cf(x) \!<\! 1 - \cf(y)$ due to $\cf(0)\!=\!0$ and continuity. 
Hence, all schemes $\efN$ based on $\cf \ne \cf_{(u)}$ fail to satisfy 
\eqref{eq:ssp}, and $\Nmapssup$ is spanned by $\efN_{(u)}$. 

\smallskip

\noindent
{\bf 5. Uniqueness of ECD.} It is now straightforward to show \eqref{eq:100}. 
Consider a pair $\efNm$, $\efN_{(u)}$ for arbitrary but fixed $0 \!<\! u \!\le\! 1$. 
We compare their counting functions $\cfu(\w)$ and $\cf_{(u)}(\w)$ on the following
partition of their domain. 
$(i)$ $0 \le \w \le u$. Here $\cfu(\w) \!=\! \w$ and $\cf_{(u)}(\w)=c/u$ and hence 
$\cf_{(u)}(\w) = \cfu(\w)/u$.
$(ii)$ $u < c < 1$. Here $\cfu(\w)=\w$ and $\cf_{(u)}(\w)=1$ and so
$\cf_{(u)}(\w)=\cfu(\w)/\w < \cfu(\w)/u$.
$(iii)$ $c \ge 1$. Here $\cfu(\w)=\cf_{(u)}(\w)=1$ and so $\cf_{(u)}(\w) \le \cfu(\w)/u$.

Taken together, this yields $\cfu(c) \le \cf_{(u)}(\w) \le \cfu(\w) / u$ for any 
$\w \!\ge\! 0$, as well as 
\begin{equation}
    \efNm[\W] \le \efN_{(u)}[\W] \le \frac{1}{u}\, \efNm[\W]
    \;\; ,\, \forall \, \W
    \; ,\; \forall \, u \!\in\! (0,1]
    \label{eq:240}
\end{equation}
Now, consider a regularization sequence $\{C\}$ such that its ECD \eqref{eq:060} 
associated with $\efNm$ (i.e. $\Delta_\star$) exists. From \eqref{eq:240} it
follows that the power governing the growth of $\efN_{(u)}[\W_k]$ with 
$\nrN_k$ also exists and is equal to $\Delta_\star$ as claimed
in \eqref{eq:100}. Hence, the concept of effective counting dimension 
associated with $\{C\}$ is well-defined (unique). 

\smallskip

\noindent
{\bf 6. Generality of ECD.} The general context we associated with ECD,
namely that of arbitrary collections $\Obj$ of objects, may raise questions about 
using the term ``dimension". Indeed, its intuitive notion is frequently reserved for 
less general setups and, in particular, for those involving metric (see comment 
{\em (i)} in Sec.~2). We thus elaborate more on the underlying rationale.

In line with the usual practice, we aimed at minimal conditions under which 
the notion of ECD is applicable. Such minimal setup turns out to be a sequence 
$(O_k,C_k)$ of objects and associated additive weights. This arises due to 
the fact that, while ordinary counting doesn't give any structure to this most bare 
of settings, the effective counting does. In fact, the most relevant consequence 
of present analysis is that ECD provides for the robust and well-founded 
quantitative characteristic of this structure.

To illustrate the reasoning, consider an extreme example of a sequence 
where $(O_1,C_1)$ involves apples, $(O_2,C_2)$ potatoes, $(O_3,C_3)$ apples 
again, $(O_4,C_4)$ peanuts and so on. While a casual observer presented with 
the sequence may be puzzled by its meaning, it may have a clear rationale for 
a fertilizer company which generated it as part of their efficiency analysis. 
The associated ECD has the same nominal meaning for both (it specifies how 
effective number of objects scales with their ordinary number), but the fertilizer
company will find it natural to call it dimension.  After all, at the heart of ordinary 
measure-based dimensions is the scaling of measure, and they are dealing with
scaling of their own measure represented by the effective count. The casual observer, 
from whom the meaning of effective count is hidden, may object. 

In the next section we will discuss an example of ECD stability study using critical wave 
functions of 3d Anderson transitions. This involves sequences labeled by size 
$L$ of the system which, similarly to the example above, may seem like sequences of 
independent distinct objects. However, they have common origin in probability 
distributions of wave functions generated by quantum dynamics of the Anderson 
model. This makes the ensuing sequences meaningful. Moreover, the objects are actually 
elements of physical space in this case, which allows us to interpret ECD 
as an effective dimension of space occupied by the Anderson electron.

\smallskip

\noindent
{\bf 7. Anderson Criticality.}  
The results of Ref.~\cite{Horvath:2018aap} and~the present work suggest 
that using $\efNm$ alone suffices for many effective counting analyses. 
But even then, additional input from other schemes in $\Nmaps$ may be 
informative. 
For example, it can be used to assess the reliability of regularization removals.
Indeed, $\Nmapssup$ is spanned by schemes $\efN_{(u)}$ and we showed 
that the associated $\Delta(u)$ is constant. However, carrying out the 
$k \!\to\! \infty$ extrapolation in practice can be affected by large systematic 
errors if available collections are not sufficiently large to achieve scaling 
in~\eqref{eq:060}. The computed $\Delta(u)$ is then expected to vary 
significantly. On the other hand, when approximate scaling is in place, 
the degree of non-constant behavior can be used to judge the level of 
systematic errors. 

To explain, note first that possible effective supports $\Obj_s \!=\! \Obj_s(u)$ of $\Obj$ 
contain populations of $\efN_{(u)}$ objects that decrease with increasing $u$. 
Hence, for given $\cO \!=\! \{\Obj_{k_1}, \Obj_{k_2}, \ldots,\Obj_{k_M}\}$ used 
in regularization removal, function $\Delta(u,\cO)$ is also expected to be 
decreasing. At the same time, $u$ can be lowered to make the fraction of objects 
in effective support arbitrarily close to one and there is a guaranteed 
over-representation of the scaling population at sufficiently small $u$. 
In fact, it is expected that $\lim_{u\to 0} \Delta(u,\cO) \!\approx\! 1$ for 
generic $\cO$, regardless of true ECD. The signature of $\cO$ suitable 
for regularization removal is the existence of a ``scaling window" in $u$, 
where $\Delta(u,\cO)$ changes slowly. The value $\Delta(u_0,\cO)$ at point 
$u_0$ of slowest change is expected to produce the most reliable estimate 
of ECD from $\cO$. The change of $\Delta(u,\cO)$ within the window sets 
an approximate scale of systematic error.

We now apply this general strategy to the recent 
calculation~\cite{Horvath:2021zjk} of spatial effective dimensions $d_\fir$ 
at Anderson transitions~\cite{Anderson:1958a,Evers_2008,MacKinnon:1981a} 
in three dimensions ($d_\fir \!=\!3\Delta_\fir$).
We will focus on 3d Anderson model in the orthogonal class, defined
on $(L/a)^3$ cubic lattice with sites labeled by $r\!=\!(x_1,x_2,x_3)$ and 
periodic boundary conditions. The model is diagonal in spin and it is 
thus sufficient to consider 1-component fermionic operators $c_r$. 
Denoting by $\epsilon_r$ the on-site random energies chosen from 
a box distribution in the range $[-W/2,+W/2]$, 
the Hamiltonian~is~\cite{Anderson:1958a}
\begin{equation}
     {\cal H} \,=\, \sum_r \epsilon_r \, c^\dag_r  \, c_r 
     \,+\, \sum_{r,j}  c^\dag_r \, c_{r-e_j} + h.c.
     \label{eq:250}
\end{equation}
Here $e_j$ ($j\!=\!1,2,3$) are unit lattice vectors. For energy $E\!=\!0$,
there is a critical point at $W \!=\! W_c\!=\!16.543(2)$~\cite{Slevin_2018} 
separating extended states at $W \!<\! W_c$ from exponentially localized 
ones at $W \!>\! W_c$. 

Objects $o_i$ involved in the calculation of $d_\fir \!=\! d_\fir(E,W)$ are 
elementary cubes of space at positions $r_i$, with weights specified by 
wave function $\psi$ via $\wl_i \!=\! p_i \!=\! \psi^+\psi(r_i)$. Collection $\Obj$ 
forms the space occupied by the system with volume 
$V \!=\! \nrN[\Obj] a^3 \!=\! L^3$. 
Electron in state $\psi$ is effectively present in a subregion $\Obj_\star[\psi]$ 
of volume $V_{\text{eff}} \!=\! \efNm[\psi] a^3$. Dimension $d_\fir$ gauges 
the asymptotic response of $V_{\text{eff}}$ to increasing $L$. The model 
involves averaging over disorder $\{\epsilon_r\}$, and hence
$\efNm \!\rightarrow\! \langle \,\efNm \rangle$ in definition~\eqref{eq:110}. 
The critical effective dimension $d_\fir(0,W_c) \!\approx\! 8/3$ was found for
\eqref{eq:250} and models in three other universality 
classes~\cite{Horvath:2021zjk}. This commonality was expressed by 
super-universal value $d_\fir^{\text{su}} \!=\! 2.665(3)$ with the quoted 
uncertainty including the spread over classes.

\begin{figure}[t]
   \includegraphics[width=0.40\textwidth]{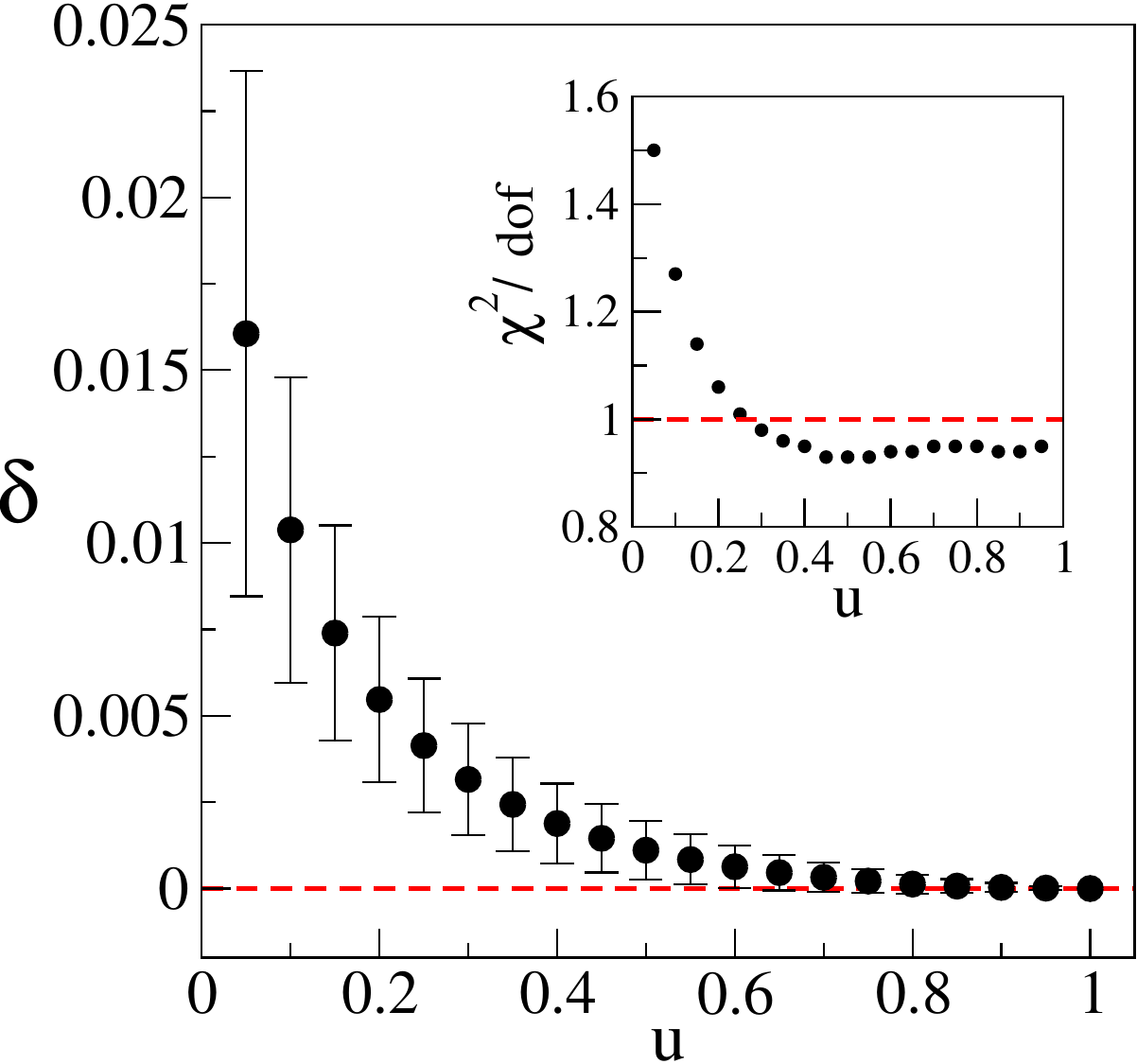}
   \caption{Deviation $\delta(u,\cO) \!=\! d_\fir(u,\cO) \!-\! d_\fir(1,\cO)$ obtained 
   via 2-power fit as described in the text. The inset shows $\chi^2$ per degree
   of freedom (dof) for the fits involved.}
    \label{fig:delta}
    \vskip -0.17in
\end{figure}

We have generated a new set of data for system~\eqref{eq:250} at the critical 
point $(0,W_c)$, producing $\cO$ including 26 systems with sizes in the range 
$22 \!\le\! L/a \!\le\! 160$. JADAMILU library~\cite{jadamilu_2007} was used 
for matrix diagonalization. We then strictly followed the analysis procedure of 
Ref.~\cite{Horvath:2021zjk} and obtained $d_\fir(\cO)=2.6654(11)$, consistently 
with the original estimate. However, in the present calculation 
of critical eigenmodes we recorded $\efN_{(u)}$ for 20 equidistant values of 
$u$ starting with $u\!=\! 0.05$, rather than just $\efNm \!=\! \efN_{(1)}$. This 
allows us to perform the proposed stability analysis utilizing $\Nmapssup$.

The latter is most efficiently performed by computing 
$\delta(u,\cO) \!=\!d_\fir(u,\cO) \!-\! d_\fir(1,\cO)$ which can
be extracted directly from the raw data without any intermediate steps.
Indeed, the large $L$ behavior of
$\langle \efN_{(u)} \rangle_L/  \langle \efN_{(1)} \rangle_L$ is governed 
by power $\delta(u)$. Moreover, relation $d_\fir(u) \!=\! d_\fir$ is replaced 
by the definite $\delta(u) \!=\! 0$. While still featuring the expected 
decreasing behavior and slow variation in the scaling window, 
the size of $\delta(u,\cO)$ directly conveys the scale of systematic errors. 
Note that since the above ratio defining $\delta(u)$ involves correlated
data in the way we performed the calculation, Jackknife procedure was used
to estimate its error in the analysis described below.

To extract $\delta(u,\cO)$ from the data in an unbiased way, we included 
it as a parameter in general 2-power fits of the above ratio in the form 
$c_1 L^\delta + c_2 L^{-\beta}$. The role of the second power is to absorb 
finite-volume effects, and its presence resulted in very stable results. 
Unconstrained 2-power fits were mainly afforded by our extensive statistics 
(30K--100K of disorder realizations).
We proceeded by finding the smallest size $L_{\text{min}}$ in $\cO$, such 
that the fit in the range $L_{\text{min}}/a \!\le\! L/a \!\le\! 160$ yielded
$\chi^2/\text{dof} \!<\!1$ for $u\!=\!0.95$ data. The resulting 
$L_{\text{min}}=30a$ was then fixed for fits at all $u$, leading 
to $\delta(u,\cO)$ shown in Fig.~\ref{fig:delta}. The respective 
$\chi^2/\text{dof}$ are shown in the inset.

The resulting $\delta(u,\cO)$ is indicative of $\cO$ suitable for regularization 
removal. Indeed, populations associated with the window $0.75 \!\le\! u \!\le\! 1$ 
scale essentially in sync. 
The slowest change occurs at $u_0 \!=\!1$, suggesting that the quoted 
result for $d_\fir$, which is based on $\efNm$, is nominally the most reliable 
for this $\cO$. Note that, according to Eq.~\eqref{eq:240}, effective support 
at $u\!=\!1/2$ is up to twice as abundant as the minimal one. The associated 
$\delta(1/2,\cO) \!\approx\! 0.002$ offers a convenient canonical benchmark  
for the level of systematic error. Given the position of the scaling 
window and its degree of stability, it is likely an upper bound in this 
case. These findings suggest that $\approx \! 10^{-3}$ is the scale of statistical 
as well as systematic error associated with calculation of $d_\fir$
in Ref.~\cite{Horvath:2021zjk}.

\begin{acknowledgments}
   P.M. was supported by Slovak Grant Agency VEGA, Project n. 1/0101/20.
\end{acknowledgments}

\bibliography{my-references}

\end{document}